\documentclass[superscriptaddress,twocolumn,prb,showpacs,preprintnumbers,amsmath,amssymb]{revtex4}
\usepackage{graphicx}
\begin{document}

\title{Magnetic field-enhanced spin freezing in YBa$_2$Cu$_3$O$_{6.45}$ at the verge of the competition between superconductivity and charge order}

\author{T. Wu}
\affiliation{Laboratoire National des Champs Magn\'etiques Intenses, UPR 3228, CNRS-UJF-UPS-INSA, BP166, F-38042 Grenoble Cedex 9, France}
\author{H. Mayaffre}
\affiliation{Laboratoire National des Champs Magn\'etiques Intenses, UPR 3228, CNRS-UJF-UPS-INSA, BP166, F-38042 Grenoble Cedex 9, France}
\author{S. Kr\"amer}
\affiliation{Laboratoire National des Champs Magn\'etiques Intenses, UPR 3228, CNRS-UJF-UPS-INSA, BP166, F-38042 Grenoble Cedex 9, France}
\author{M. Horvati\'c}
\affiliation{Laboratoire National des Champs Magn\'etiques Intenses, UPR 3228, CNRS-UJF-UPS-INSA, BP166, F-38042 Grenoble Cedex 9, France}
\author{C. Berthier}
\affiliation{Laboratoire National des Champs Magn\'etiques Intenses, UPR 3228, CNRS-UJF-UPS-INSA, BP166, F-38042 Grenoble Cedex 9, France}
\author{C.T.~Lin}
\affiliation{Max-Planck-Institut for Solid State Research, Heisenbergstra{\ss}e 1, D-70569 Stuttgart, Germany} %
\author{D. Haug}
\affiliation{Max-Planck-Institut for Solid State Research, Heisenbergstra{\ss}e 1, D-70569 Stuttgart, Germany} 
\author{T. Loew}
\affiliation{Max-Planck-Institut for Solid State Research, Heisenbergstra{\ss}e 1, D-70569 Stuttgart, Germany} 
\author{V. Hinkov}
\affiliation{Max-Planck-Institut for Solid State Research, Heisenbergstra{\ss}e 1, D-70569 Stuttgart, Germany} 
\affiliation{Quantum Matter Institute, University of British Columbia, Vancouver V6T0A5, Canada}
\author{B. Keimer}
\affiliation{Max-Planck-Institut for Solid State Research, Heisenbergstra{\ss}e 1, D-70569 Stuttgart, Germany} 
\author{M.-H. Julien}
\email{marc-henri.julien@lncmi.cnrs.fr}
\affiliation{Laboratoire National des Champs Magn\'etiques Intenses, UPR 3228, CNRS-UJF-UPS-INSA, BP166, F-38042 Grenoble Cedex 9, France}

\pacs{74.25.nj, 74.25.Ha, 74.72.Kf}

\begin{abstract}

Using $^{63}$Cu NMR, we establish that the enhancement of spin order by a magnetic field $H$ in YBa$_2$Cu$_3$O$_{6.45}$ arises from a competition with superconductivity because the effect occurs for $H$ perpendicular, but not parallel, to the CuO$_2$ planes, and it persists up to field values comparable to $H_{c2}$. We also find that the spin-freezing has a glassy nature and that the frozen state onsets at a temperature which is independent of the magnitude of $H$. These results, together with the presence of a competing charge-ordering instability at nearby doping levels, are strikingly parallel to those previously obtained in La-214. This suggests a universal interpretation of magnetic field effects in underdoped cuprates where the enhancement of spin order by the field may not be the primary phenomenon but rather a byproduct of the competition between superconductivity and charge order. Low-energy spin fluctuations are manifested up to relatively high temperatures where they partially mask the signature of the pseudogap in $1/T_1$ data of planar Cu sites.

\end{abstract}
\maketitle

\section{Introduction}

One of the most debated issue in high $T_c$ cuprates is whether the pseudogap, and perhaps superconductivity itself, are related to the occurrence of electronic ordering competing with superconductivity~\cite{competition}. Why has this issue not been settled yet? First, despite a plethora of ordering phenomena reported in these materials, there are actually very few instances where a competition with superconductivity is unambiguously demonstrated. The strongest evidence comes from those experiments which have shown either 1) that a magnetic field promotes spin and/or charge ordering if and only if this field weakens significantly superconductivity~\cite{Hoffman02,Lake02,Lake05,Wu11,Wu13}, or 2) that the occurrence of one order reduces the amplitude of its competitor~\cite{Tranquada95,Ghiringhelli12,Achkar12,Chang12b}. Second, it has been unclear whether there is a single, generic, competing order behind the different phenomena observed. In particular, which of spin order or charge order would be the leading phenomenon in that case has been a central question~\cite{Arovas97,Sachdev02b,Kivelson02b,Sachdev09}. These issues obviously need to be settled before discussing any possible relation to the mechanism of superconductivity.

\begin{figure}[b!]
\label{interaction}
\centerline{\includegraphics[width=3.3in]{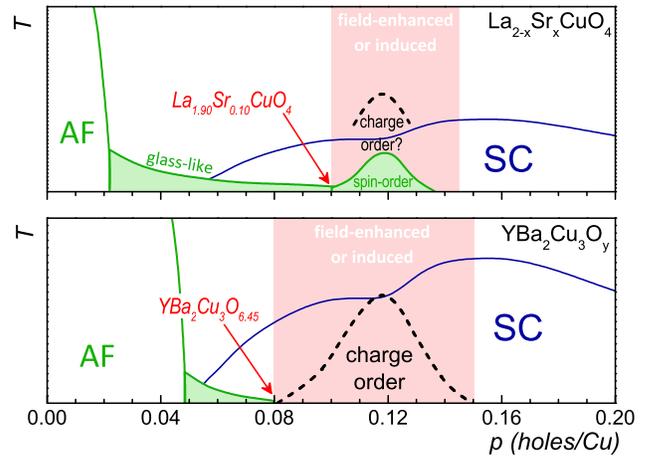}} 
\vspace{-0.2cm} \caption{(Color online) Phase diagrams of La$_{2-x}$Sr$_x$CuO$_4$ (a) and YBa$_2$Cu$_3$O$_{y}$ (b), based on data in refs.~\cite{Julien03,Wu11,Wu13,LeBoeuf11,Laliberte11,Coneri10}. The shaded regions are those where field-tuned competing orders have been reported. The dotted line in (a) represents a putative charge-ordering transition which, although not directly observed, is suspected to be present by analogy with La$_{2-x}$Ba$_x$CuO$_4$ and La$_{2-y-x}$Eu/Nd$_y$Sr$_x$CuO$_4$~\cite{Vojta09}. The boundary of the (field-induced) charge-ordered state in (b) is based on an interpretation of transport measurements~\cite{LeBoeuf11,Laliberte11,Wu11}.}
\end{figure}

Recently, an important result was obtained in YBa$_2$Cu$_3$O$_{6.45}$~\cite{Haug09,Haug10} where an enhancement of spin order by a magnetic field was observed for the first time in a cuprate other than La-214~\cite{Lake02,Lake05,Katano00,Khaykovich05,Chang08,Fujita06,Savici05,Machtoub05,Wen12}. A significant step for establishing the ubiquity of competing orders in cuprates would be to demonstrate that the field-induced magnetism in YBa$_2$Cu$_3$O$_{6.45}$ arises from the weakening of superconductivity by the field, rather than from a direct effect of the field itself~\cite{Haug10,Stock09}. More generally, the degree of similarity between these two different families (Fig.~1) is related to the central, but controversial, issue of universality in cuprates~\cite{Haug10,Hinkov08,Stock09,Stock06,theo6.45}.
\begin{figure*}[t!]
\label{spectrum}
\centerline{\includegraphics[width=7in]{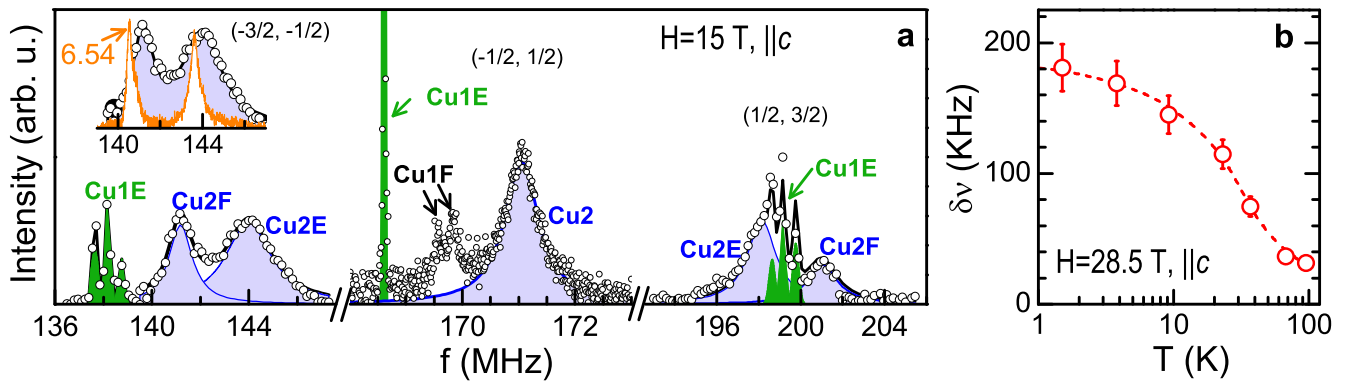}} 
\vspace{-0.2cm}
\caption{(Color online) (a) $^{63}$Cu NMR spectrum ($T$=60~K) with quadrupole "satellites" ($\pm3/2,\pm1/2$ transitions) and central lines (+1/2,-1/2). The line assignment is made on the basis of previous NMR works in YBa$_2$Cu$_3$O$_y$. Note that "Cu2E" is an approximate naming here because, when the oxygen content is lower than y=6.5, this line most likely also contains those Cu2F which lie below chain-oxygen vacancies~\cite{Wu}. This explains why the width and integrated area of this Cu2E line are larger than for Cu2F. Left inset: comparison with Cu2 satellites of YBa$_2$Cu$_3$O$_{6.54}$~\cite{Wu11}. (b) Full-width-at-half-maximum $\delta\nu$ (dashes are a guide to the eye) of the Cu1E central line. The shift of this line is negligible with respect to $\delta \nu$. }
\end{figure*}

Here we report $^{63}$Cu nuclear magnetic resonance (NMR) results in YBa$_2$Cu$_3$O$_{6.45}$ untwinned single crystals from the same batch as those studied in neutron scattering~\cite{Haug09,Haug10}. Owing to the unique capability of NMR to be operated in any field orientation and in the highest achievable field strengths, our results demonstrate that the field indeed promotes a competing order at the expense of superconductivity, as it does in La-214. We further uncover a number of other similarities with La-214 and we make the, previously unnoticed, observation that {\it all} of the existing evidence of field-induced effects in underdoped cuprates has been obtained in a specific doping range over which superconductivity competes with a charge ordering instability (Fig.~1). This leads us to suggest a universal interpretation of magnetic field effects in underdoped cuprates where the enhancement of spin order by the field is not the central phenomenon, but rather a byproduct of the competition between superconductivity and charge order.

\section{Superconducting properties}

A sharp superconducting transition is observed in zero field at $T_c=35$~K. In a field, the melting temperature of the vortex lattice $T_{\rm melt}$ was measured through the resonance frequency of the NMR tank circuit and the obtained values (Fig.~3d) were in agreement with data in YBa$_2$Cu$_3$O$_y$ at a similar doping level~\cite{Ramshaw12}. $T_{\rm melt}$ decreases much more with increasing fields $H\|c$ than with $H\|ab$, as expected for this anisotropic superconductor (Fig.~3d). It is important to realize that the relaxation peak reported below cannot be related to the solid to liquid transition of the vortex lattice: the central line broadening starts far above $T_c$. At 28.5~T, it has reached 90~\% of its low $T$ value before $T_{\rm melt}$ and it corresponds to an amplitude of field distribution which is about three time larger than that possibly due to the vortex lattice. The peak in $1/T_1$ manifestly corresponds to the spin freezing seen by neutron scattering or muon spin rotation ($\mu$SR) and the peak temperature is unrelated to $T_{\rm melt}$ which shows an opposite field dependence.

\section{C\lowercase{u}2 and C\lowercase{u}1 NMR spectra}

The presence of two distinct sites in the NMR spectra of planar Cu2 (Fig.~2a) reveals an intrinsic degree of ortho-II oxygen order in the chains~\cite{Andersen99}. Nonetheless, disorder related to oxygen vacancies in the oxygen-filled chains is observed, as expected from the actual doping level $y\simeq 6.45$ (hole content $p\simeq0.07-0.08$) being lower than the ideal ortho-II composition $y=6.50$: there are three different Cu1E (empty chains) sites and two Cu1F (oxygen-filled chains) sites instead of one Cu1E and one Cu1F for $y=6.50$. These multiple sites correspond to the different positions with respect to the oxygen vacancies~\cite{Wu}.

The signal from Cu2 sites can be studied from room temperature down to $\sim$50~K and we show in the Appendix that such measurements provide evidence for low-energy magnetic fluctuations masking the signature of the pseudogap in the spin-lattice relaxation rate (1/$T_1$) data of Cu2 so that no decrease of the magnitude of the pseudogap should be deduced from these measurements. Below 50~K, however, the only reliable NMR signal is that from $^{63}$Cu1E sites which, unlike planar Cu2, is not wiped out by too fast relaxation times as the magnetic transition is approached (see Appendix).

The strong broadening (Fig.~2b) of the Cu1E line shows that a distribution of hyperfine fields $\langle h_z \rangle =g A\langle S_z \rangle$ develops on cooling in the CuO$_2$ planes, most likely due to oxygen disorder in the chains impacting onto the planes~\cite{Chen09} ($A$=0.3~T/$\mu_B$~\cite{Dooglav96} is the hyperfine coupling to the CuO$_2$ planes and $g$ is Land\'e's factor). At 1.5~K and 28.5~T, the value of $\delta \nu$ translates into a distribution of spin polarization of $\delta\langle S_z \rangle \simeq\pm3\times10^{-2}$~$\mu_B$.

\begin{figure*}[t!]
\label{relaxation}
\centerline{\includegraphics[width=7in]{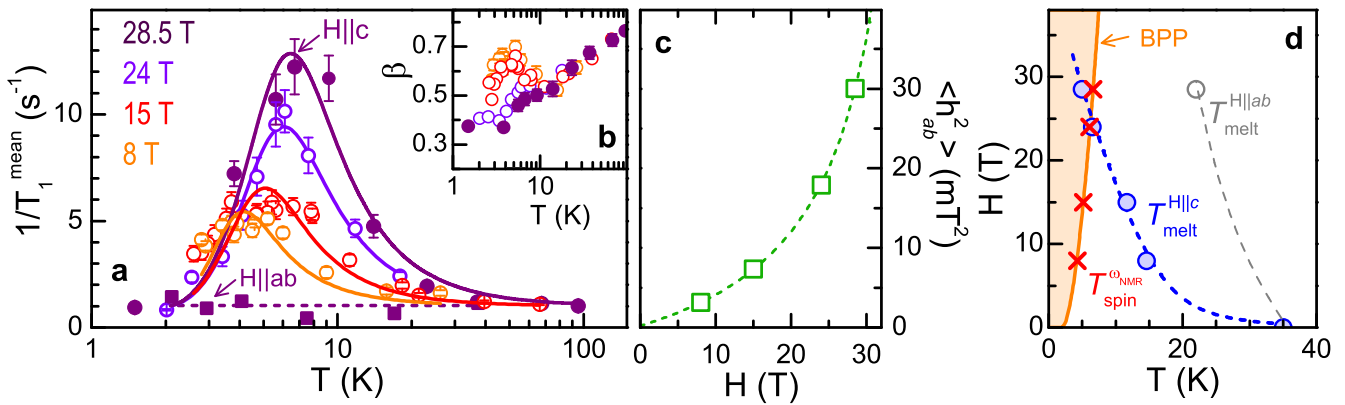}} 
\vspace{-0.2cm} \caption{(Color online) (a) 1/$T_1^{\rm mean}$ of Cu1E for $H\|c$ (except open circles for $H\|ab$). Lines are fits to the BPP formula assuming a field-independent $\tau_c=0.042/(T-2)^{1.7}$~\cite{NMRLa214}, where the numerical values were determined from fits to the data at 24~T. The only adjustable parameter between different fields is thus the amplitude factor $\langle h_{\perp}^2 \rangle=\langle h_{ab}^2 \rangle$. (b) Stretching exponent $\beta$ with same color code as in (a). (c) Mean-squared hyperfine field $\langle h_{\perp}^2 \rangle=\langle h_{ab}^2 \rangle$ at $T=T_{\rm spin}^{\rm \omega_{NMR}}$. (d) Spin-freezing temperature $T_{\rm spin}^{\rm \omega_{NMR}}$ deduced from the peaks of the $T_1$ fits (following by construction the field dependence of the BBP model) and vortex-melting temperature $T_{\rm melt}$ deduced from the detuning of the NMR tank circuit.}
\end{figure*}

\section{Glassy spin freezing}

The time-dependence of the Cu1E magnetization after saturating the central line (where the three sites are unresolved) was fit to $m(t)=m_\infty\{1-0.1\exp(-(t/T_1^{\rm mean})^\beta)-0.9\exp(-(6t/T_1^{\rm mean})^\beta)\}$ corresponding to a distribution of $1/T_1$ values whose width in a log scale is linear in $\beta$  between  $\beta=1$ ({\it i.e.} a homogeneous system) and $\beta\simeq0.5$~\cite{Johnston06,Mitrovic08}. $1/T_1^{\rm mean}$ is the most probable value of the distribution, and is approximately satisfying the standard definition of $T_1$, namely the time to reduce the perturbation by $1/e$, independently of the $\beta$ value. This makes the fitted $1/T_1^{\rm mean}$ values stable and mostly decoupled from $\beta$. The parametrization of the $T$- or $H$-dependence of $1/T_1^{\rm mean}$ with models developed for homogeneous situations (like the BPP model below) thus provides a valuable insight into the intrinsic, {\it most probable}, behavior of the system, in particular as much as the corresponding $\beta$ variations are not substantial over the relevant range (while those of $1/T_1^{\rm mean}$ are).

The increase of $1/T_1^{\rm mean}$ is evidence that the spin fluctuations slow down on cooling ({\it i.e.} their correlation time $\tau_c$ diverges), as expected from earlier evidence of spin ordering~\cite{Haug09,Haug10,Hinkov08}. In agreement with the above-noted distribution of $\langle S_z \rangle$, the decrease of $\beta$ (Fig.3a,b) shows that heterogeneity develops on cooling along with the slowing down. The possible origin of this glassy dynamics has been addressed by many authors (\cite{Haug10,Mitrovic08,Stock06,Sega09,Kivelson02a} and refs. therein). This discussion is beyond the scope of the present work but the important point here is that the very same glassy freezing is observed in La-214 cuprates.

The peak of $1/T_1^{\rm mean}$, observed here at $T_{\rm spin}^{\rm \omega_{NMR}}\simeq4-6$~K, is expected when the characteristic frequency of spin fluctuations becomes as small as the NMR frequency $\omega_{\rm NMR}$. As in previous studies of La-214~\cite{NMRLa214}, the $T$ dependence of $1/T_1^{\rm mean}$ can be accounted for by the "BPP" formula~\cite{Slichter}
${1 \over T_1} = \gamma_n^2 \left< h_{\bot}^2 \right> {2 \tau_c \over {1 + \omega_{\rm NMR}^2 \tau_c^2}}$,
where $h_\bot$ is the component of the hyperfine field perpendicular to $H$ and $\gamma_n$ is the nuclear gyromagnetic ratio. Since we are analyzing $1/T_1^{\rm mean}$ here, $\tau_c$ represents the most probable correlation time of the fluctuating spin system.

\section{Probe-frequency dependence}

The values of $T_{\rm spin}^{\rm \omega_{NMR}}=4-6$~K are much lower than $T_{\rm spin}^{\rm NS}\simeq30$~K, the temperature at which an elastic signal appears in neutron scattering. As shown by numerous studies of glasses, it is the broad temperature range of inhomogeneous slowing down which makes the apparent freezing temperature $T_{\rm spin}$ probe-frequency dependent. The difference of temperatures becomes significant here because the timescale of the neutron scattering experiment (100~$\mu$eV resolution here) is hundred times faster than the typical NMR timescale of 10$^{-2}\mu$s. This shows that the closeness of the values of $T_{\rm spin}^{\rm NS}\simeq$30~K and $T_c=35$~K is essentially fortuitous: spin order does not occur near $T_c$ but at much lower temperature. Moreover, $\mu$SR experiments with a timescale of 10~$\mu$s (thus slower than NMR) detect static fields below $T_{\rm spin}^{\mu}\simeq 2$~K in this sample~\cite{Hinkov08}, which is logically lower than $T_{\rm spin}^{\rm \omega_{NMR}}=4-6$~K and is close to $T_{\rm spin}^{\omega=0}=2\pm2$~K, the freezing temperature extrapolated down to zero frequency with the BPP formula (Fig.~3d). This probe-frequency dependence of $T_{\rm spin}$ is another striking similarity with the La-214 materials.

\section{Evidence of competition with superconductivity}

As Fig.~3a shows, the amplitude of $1/T_1$ depends strongly on field up to (at least) 28.5~T for $H\|c$. Since $(1/T_1)^{\rm max}=\gamma_n^2 \left< h_{\bot}^2 \right>/\omega_{\rm NMR}$ at $T=T_{\rm spin}^{\rm \omega_{NMR}}$, the mean-squared fluctuating hyperfine field $\left< h_{\bot}^2 \right>$ can be extracted at this temperature without supposing any $T$ dependence for $\tau_c$. It appears that $\left< h_{\bot}^2 \right>$ strongly varies with field (Fig.~3b). Furthermore, whatever the supposed form of $\tau_c(T)$, fits in Fig.~3a show that the $1/T_1$ data are well described by a BPP model without any field dependence of $\tau_c$. Of course, the fits could be improved (especially at low fields) with a field-dependent $\tau_c$ but this is a minor correction which does not affect the results of this paper and the main message here is that field dependence of $1/T_1$ is, to a first approximation, entirely explained by the field dependence of $\left< h_{ab}^2 \right>$. This field-dependence should naturally be related to the observed enhancement of $I_{\rm NS}$ by the field~\cite{Haug09,Haug10}. Moreover, a similar field-enhancement of $T_1$ is observed at low $T$ in La$_{1.90}$Sr$_{0.10}$CuO$_4$~\cite{unpubl} for which the quasi-elastic neutron scattering intensity $I_{\rm NS}$ (in general proportional to the square of the ordered moment) is also field dependent~\cite{Lake02}.

The most direct evidence for competing orders, and therefore the most important result of this work, is the absence of a peak in $1/T_1$ when the field of 28.5~T is applied parallel to the CuO$_2$ planes. Since the peak must exist for any field orientation (frozen moments of $\sim$~0.1~$\mu_B$ are detected in $\mu$SRbelow 2~K, already for $H=0$~\cite{Hinkov08,Coneri10}), $\left< h_{\bot}^2 \right>$ has to be smaller by more than an order of magnitude for $H\|ab$ than for $H\|c$ in order to make $1/T_1$ so small that no peak is seen above the relaxation background. The striking difference between $H\|c$ and $H\|ab$ cannot be explained by an extreme anisotropy of magnetic fluctuations $\left< h_{c}^2 \right>\ll\left< h_{ab}^2 \right>$ because $\left< h_{ab}^2 \right>$ also contributes to $\left< h_{\bot}^2 \right>=\frac{1}{2}\left(\left< h_{c}^2 \right>+\left< h_{ab}^2 \right>\right)$ when $H\|ab$. Instead, this result demonstrates that the enhancement of $1/T_1$ is not related to the field itself but it is due to the weakening of superconductivity which is indeed much more efficient for $H\|c$ than for $H\|ab$ (Fig.~3d). This is precisely the phenomenology expected for competing orders. In this scenario, no field dependence should exist above the upper critical field $H_{\rm c2}(\|c)\simeq40\pm5$~T for this doping level~\cite{Ramshaw12,Ando02b,Chang12}. Although this limit could not be reached here, our measurements probing the spin order for the first time in very high fields show that the enhancement of magnetism seems to occur on a field scale similar to that for the destruction of superconductivity, with $\left< h_{ab}^2 \right>$ varying as $a_1+a_2(H/(H_c-H))$ and $H_c\simeq39\pm1$~T (the use of this dependence has no other justification than introducing a field-scale $H_c$).

Studies in La$_{2-x}$Sr$_x$CuO$_4$~\cite{Machtoub05,Savici05,Sonier07} and YBa$_2$Cu$_3$O$_y$~\cite{Sonier07} have suggested that the field tunes the relative volume fractions of the superconducting and magnetic phases, consistent with the idea that the order competing with superconductivity is enhanced in and about the vortex cores. This two-phase description might also rationalize the counterintuitive observation that the neutron scattering intensity in La$_{1.90}$Sr$_{0.10}$CuO$_4$ has a field-independent onset despite its field-dependent amplitude~\cite{Lake02}. Our data in YBa$_2$Cu$_3$O$_{6.45}$ also support a field-independent $T_{\rm spin}$. Indeed, in the BPP description, the peak temperature $T_{\rm spin}^{\rm \omega_{NMR}}$ is an apparent freezing temperature which depends on $\omega_{\rm NMR}$. In a standard NMR experiment, however, $\omega_{\rm NMR}$ increases linearly with field. Therefore, $T_{\rm spin}^{\rm \omega_{NMR}}$ necessarily depends on field in NMR, although it would be field independent for any experimental probe operating at a fixed frequency $\omega_0$. The excellent fit to the peak positions in Fig.~3a (with no field dependent parameter other than the amplitude $\left< h_{ab}^2 \right>$) demonstrates that $T_{\rm spin}^{\rm \omega_{NMR}}$ values follow the $H$ dependence expected from the trivial frequency dependence of the BPP model (Fig.~3d). This implies that $T_{\rm spin}^{\rm \omega_{NMR}}$ would indeed be field-independent if $H$ could be varied without changing $\omega_{\rm NMR}$.

\section{Discussion}

In addition to demonstrating the existence of competing orders, our results have revealed striking analogies between YBa$_2$Cu$_3$O$_{6.45}$ and La-214 materials: similar glassy spin freezing, field-independent onset $T_{\rm spin}$, enhancement of magnetism for $H\|c$.

In La$_{1.90}$Sr$_{0.10}$CuO$_4$, it is generally admitted~\cite{Sachdev02b,Kivelson02b,Wen12} that the enhancement of spin order by the field actually occurs because incipient charge order~\cite{Wu12} competes with superconductivity (and is thus field-dependent). Namely, it is this charge order which, in turn, boosts spin order. This explanation does not suppose anything regarding the origin of spin order in zero field, which could well be triggered by quenched disorder rather than by charge order~\cite{Lake02,Sasagawa02,Andersen11}.

As Fig.~1 shows, despite its different doping level, YBa$_2$Cu$_3$O$_{6.45}$ ($p\simeq0.075$) lies in a similar position in the phase diagram as La$_{1.90}$Sr$_{0.10}$CuO$_4$, {\it i.e.} at the border between the glass-like phase and the region of field-dependent charge order. This leads us to propose the same description in both compounds: regardless of the (debated) origin of the (incommensurate) spin order in zero-field~\cite{theo6.45} for YBa$_2$Cu$_3$O$_{6.45}$, its enhancement by a field does not necessarily imply a mere competition between superconductivity and spin order. Instead, the field-dependent spin order could be a more subtle byproduct of the recently-discovered competition between superconductivity and a charge ordering instability at nearby doping levels~\cite{Wu11,Wu13,Laliberte11,Ghiringhelli12,Achkar12,Chang12b}. Of course, a direct evidence of charge-density-wave correlations in YBa$_2$Cu$_3$O$_{6.45}$ would be desirable but, as far as Cu NMR is concerned, disorder arising from the CuO chains and Cu(2) signal wipeout preclude the observation of a quadrupole splitting or broadening of the lines which would indicate a static charge density modulation in this sample. From this NMR point of view, the situation recalls more La-214 than YBa$_2$Cu$_3$O$_{6.6\pm0.1}$. Still, there is a real possibility that a charge instability persists down to this doping level~\cite{Ghiringhelli12,Hinkov08,LeBoeuf11,Ando02}, although it may be difficult to detect because higher oxygen disorder and stronger scattering from spin fluctuations could result in lower intensity and shorter correlation length.

In a broader perspective, the following, so far unnoticed, observation can be made: to the best of our knowledge, {\it all} of the field-induced effects (shaded areas in Fig.~1) were actually observed at doping levels around $p\simeq0.12$ where a charge-ordering instability manifestly competes with superconductivity~\cite{Wu11,Wu13,Tranquada95,Ghiringhelli12,Achkar12,Chang12b}. Likewise, no effect of the field has ever been reported in the "glass-like" region of the phase diagram in Fig.~1 (see particularly ref.~\cite{Haug10} for YBa$_2$Cu$_3$O$_{6.35}$). It is therefore possible that all of the observations of field-enhanced spin order in cuprates are actually explained by a competition between superconductivity and charge order.

Note added: After completion of this manuscript, several related papers appeared. Spin freezing has been observed with NMR in YBa$_2$Cu$_3$O$_{6.45}$ but its dependence on the magnitude and on the orientation of the field was not studied~\cite{Grafe12}. Two X-ray scattering studies in YBa$_2$Cu$_3$O$_{6.54}$ (ortho-II)~\cite{Blackburn13,Blanco13} report a charge ordering wave vector unrelated to that of the spin modulation in YBa$_2$Cu$_3$O$_{6.45}$. These result prompt for further investigation of charge correlations in YBa$_2$Cu$_3$O$_{6.45}$ and of field effects on spin order. With these results and ours, the question of universality and competing orders is more than ever center stage in cuprates.

\section{Acknowledgments}

We acknowledge helpful exchanges with P. Bourges, P. Carretta, J. Chang, N. Christensen, H.-J. Grafe, R. Liang, C. Proust, Y. Sidis, J. Tranquada, B. Vignolle, M. Vojta.
This work was supported by the 7th framework programme "Transnational Access" of the European Commission, contract N$^\circ$228043 – EuroMagNET II – Integrated Activities, by P\^ole SMIng of Universit\'e J. Fourier - Grenoble and by the French Agence Nationale de la Recherche (ANR) under reference AF-12-BS04-0012-01 (Superfield).

\section{Apprendix}

\subsection{Samples and methods}

The typical size of the untwinned, high quality, single crystals was $1.7\times 1.6\times0.4$ mm$^3$ with a sharp superconducting transition (Fig.~4). NMR measurements at 24 and 28.5~T were performed in the M10 resistive magnet of the LNCMI Grenoble. All measurements for both Cu1 and Cu2 sites were performed on the $^{63}$Cu isotope. Spin-lattice relaxation times ($T_1$) were measured by saturating the central transitions.

\begin{figure}[h!]
\centerline{\includegraphics[width=2.9in]{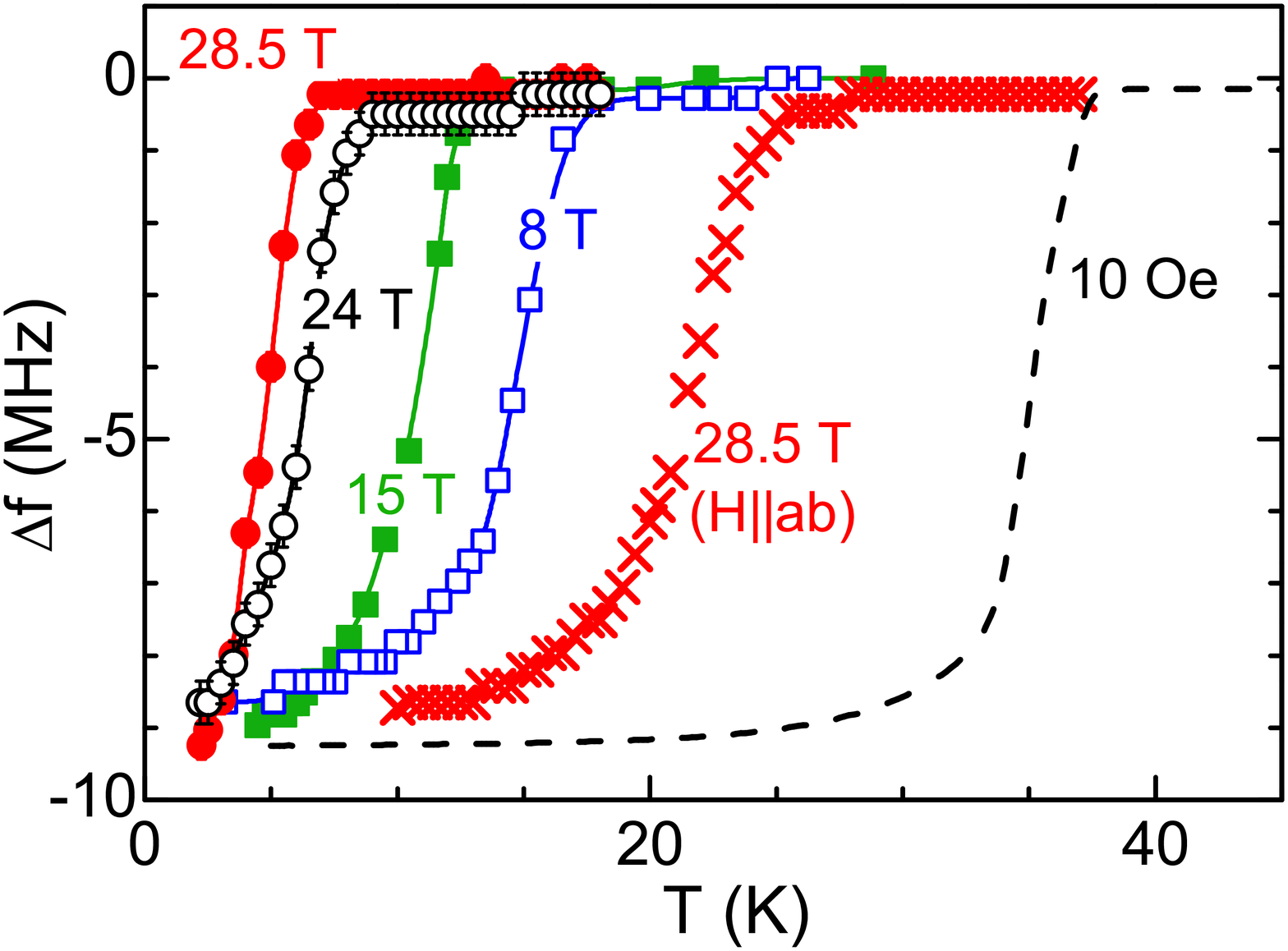}} 
\vspace{-0.2cm} \caption{Frequency shift of the resonance of the NMR tank circuit (symbols) and bulk magnetization measurements in a field of 10~Oe (dashed line, scaled to the frequency shift data). When not specified, the field is applied parallel to the $c$-axis.}
\end{figure}

\subsection{Electronic~transition at $p\simeq0.08$}
\begin{figure}[h!]
\centerline{\includegraphics[width=3in]{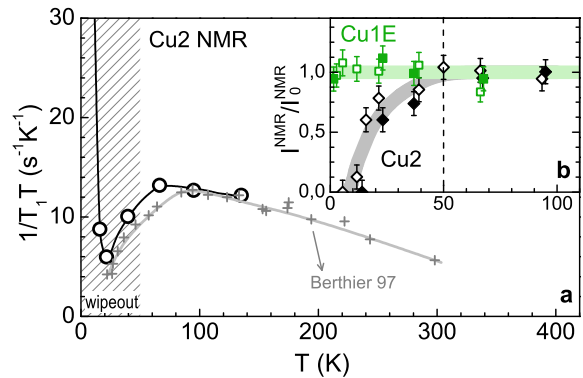}} 
\vspace{-0.2cm} \caption{(a) 1/($T_1T$) in $H\|c=15$~T, measured on the central line of $^{63}$Cu2 nuclei. The data are in agreement with an earlier study at similar doping level~\cite{Berthier97}. (b) Central-line intensity integrated over frequency and corrected for the $T_2$ decay. Unlike Cu2, the Cu1E signal does not suffer any wipeout, thus allowing reliable measurements at low $T$. Open (closed) symbols in (b) correspond to $H=15$~(28.5)~T.}
\end{figure}

1/($T_1T$) data for $^{63}$Cu2 (a measure of the dynamic spin susceptibility $\chi^{\prime\prime}(q,\omega_n)$ at the NMR frequency $\omega_n$) show a significant difference with data obtained at higher doping level in underdoped YBa$_2$Cu$_3$O$_{y}$: the values are much larger over the whole temperature range and a broad maximum occurs around $T^\dag\simeq70$~K (Fig.~5a) instead of the usual $T^\dag\simeq150$~K at higher doping~\cite{Berthier97}. We cannot exclude that no $T^\dag$ would even be observed in this sample if no wipeout occurred. Anyhow, a decrease of $T^\dag$ for $p\leq0.08$ has already been reported in YBa$_2$Cu$_3$O$_{y}$~\cite{Berthier97} as well as in 3 and 5-layer cuprates near $p\simeq0.09$~\cite{Mukuda12}. Since the maximum of 1/($T_1T$) at $T^\dag>T_c$ is typical of the pseudogap phase, it would be tempting to associate the lower $T^\dag$ to a decrease in the magnitude of the pseudogap for $p\leq 0.08$. However, $T^\dag$ is significantly smaller than $T^*$ defining the onset of the pseudogap in the uniform spin susceptibility~\cite{Alloul89} or in the resistivity~\cite{Ito93} and $T^*$ does not decrease below $y=6.5$~\cite{Alloul89,Ito93}. Also, non-magnetic impurities do not affect $T^*$~\cite{Alloul09} while they lead to a decrease of $T^\dag$ in some $T_1$ measurements~\cite{Zheng}. Actually, $k_BT^\dag$ may not represent any characteristic energy scale, especially given the complex nature of the processes involved in nuclear relaxation. Thus, the data cannot be taken as evidence of a decrease in the pseudogap energy scale below $p\simeq0.08$. Instead, the decrease of $T^\dag$ likely arises from the spin fluctuations which must be present at energy much lower than $k_B T^*$, and thus fill the pseudogap, on approaching $T_{\rm spin}$. Note that this statement does not conflict with the well-documented evidence for a crossover (or transition) from a magnetic, disorder-sensitive, 2D anisotropic, bad metal to a spin-gapped 3D metallic ground state near $p\simeq0.08$ in YBa$_2$Cu$_3$O$_{x}$~\cite{LeBoeuf11,Coneri10,Sebastian10,Rullier08,Sun04,Ando02,Li08,Baledent11,Vignolle12}. It simply means that the pseudogap may not be involved in this phenomenon.

\end{document}